\documentclass[a4paper,12pt]{article}

\usepackage{times,fullpage}
\usepackage{amsmath,amssymb,amsfonts}
\usepackage[affil-it]{authblk}
\usepackage{cite}
\usepackage{tikz}

\usetikzlibrary{positioning,arrows}
\usetikzlibrary{decorations.pathmorphing}
\usetikzlibrary{decorations.markings}

\tikzset{
fermion/.style={draw=black, postaction={decorate},
decoration={markings,mark=at position .55 with {\arrow[black]{triangle 45}}}},
vboson/.style={decorate, draw=black,decoration={snake}},
scalar/.style={dashed,draw=black}
}

\newcommand{\vvsvvS}[5]{ 
\begin{tikzpicture}[baseline=(current bounding box.center)]
\coordinate[label=above:$#1$] (e1);
\coordinate[below right=of e1] (i1);
\coordinate[below left=of i1,label=below:$#2$] (e2);
\coordinate[right=of i1] (i2);
\coordinate[above right=of i2,label=above:$#3$] (e3);
\coordinate[below right=of i2,label=below:$#4$] (e4);
\draw[vboson] (e1) -- (i1);
\draw[vboson] (i1) -- (e2);
\draw[scalar] (i1) -- node[label=above:$#5$]{} (i2);
\draw[vboson] (e4) -- (i2);
\draw[vboson] (i2) -- (e3);
\end{tikzpicture}
}
\newcommand{\vvsvvT}[5]{ 
\begin{tikzpicture}[baseline=(current bounding box.center)]
\coordinate[label=above:$#1$] (e1);
\coordinate[below right=of e1] (i1);
\coordinate[below=of i1] (i2);
\coordinate[below left=of i2,label=below:$#2$] (e2);
\coordinate[above right=of i1,label=above:$#3$] (e3);
\coordinate[below right=of i2,label=below:$#4$] (e4);
\draw[vboson] (e1) -- (i1);
\draw[vboson] (i1) -- (e3);
\draw[scalar] (i1) -- node[label=right:$#5$]{} (i2);
\draw[vboson] (e4) -- (i2);
\draw[vboson] (i2) -- (e2);
\end{tikzpicture}
}
\newcommand{\vvsvvU}[5]{ 
\begin{tikzpicture}[baseline=(current bounding box.center)]
\coordinate[label=above:$#1$] (e1);
\coordinate[below right=of e1] (i1);
\coordinate[below=of i1] (i2);
\coordinate[below left=of i2,label=below:$#2$] (e2);
\coordinate[above right=of i1,label=above:$#3$] (e3);
\coordinate[below right=of i2,label=below:$#4$] (e4);
\draw[vboson] (e1) -- (i1);
\draw[vboson] (i1) -- (e4);
\draw[scalar] (i1) -- node[label=left:$#5$]{} (i2);
\draw[vboson] (e3) -- (i2);
\draw[vboson] (i2) -- (e2);
\end{tikzpicture}
}

\begin{document}
\begin{titlepage}
	\title{Diphoton excess and VV-scattering}
	\author{ G. Cynolter\footnote{cyn@general.elte.hu}, \, 
J. Kov\'{a}cs\footnote{kovacsjucus@caesar.elte.hu} \, and 
E. Lendvai\footnote{lendvai@general.elte.hu} }
	\affil{MTA-ELTE Theoretical Physics Research Group, E\"otv\"os University,
Budapest, 1117 P\'azm\'any P\'eter s\'et\'any 1/A, Hungary}
	\date{}

    \maketitle
	\thispagestyle{empty}
    
    \begin{abstract}
		We consider minimal effective interactions of the 750 GeV mass resonance observed recently by ATLAS and CMS. Assuming a new scalar and gauge invariant effective interactions leads to non-trivial two particle scattering amplitudes with asymptotic gauge boson states.  The longitudinally polarized $W^\pm$ and $Z$ bosons interacting via dimension-five effective operators  provide stringent constraints on the validity of the effective model. The large width found by ATLAS implying a bound of  approximately 500 GeV already below the resonance, turns this scenario unlikely. For production mainly in gluon fusion we get an upper bound of $\sim 1.3$ TeV and strong limits on the masses of the  underlying vector-like fermions are given.
	\end{abstract}
\end{titlepage}

\section{Introduction}

ATLAS and CMS collaborations have recently presented excess in the diphoton searches in the 13 TeV LHC run-2 data. ATLAS \cite{ATLAS} claims events with $3.9\sigma$ local significance over the smooth background assuming spin-2 and large width (they found $3.6\sigma$ for spin-0). The re-analyzed ATLAS run-1 data presented at Moriond EW 2016 shows $2\sigma$ excess for spin-0 and it is also compatible with a spin-2 resonance. The local significance of the CMS \cite{CMS} diphoton excess grows from 2015 December to Moriond to a local $2.9\sigma$ after adding 20 percent new data to the analysis. The combined CMS run-1 and run-2 significance is $3.4\sigma$ both for spin-0 and spin-2, preferring low width. The clean diphoton signature and the absence of other final states hint towards a new scalar or pseudoscalar resonance with mass approximately 750 GeV \cite{whatistheresonance,Falkowski,Ellis:2015oso,Staub:2016dxq}. The relatively large cross section requires large couplings or strong dynamics. A spin-2 resonance could be a Kaluza-Klein graviton from extra dimensional models, but not favored by the combined run-1 and run-2 ATLAS data \cite{spin2}.

Motivated by the photon-photon final state resembling the Higgs discovery we assume that the new resonance is a new electroweak singlet scalar particle. Direct couplings to standard fermions and the Higgs must be suppressed as no fermion final state has been observed and the mixing with the Higgs is severely constrained \cite{Falkowski}. We assume that the new singlet couples to the field strength of electroweak gauge bosons and gluons via dimension five gauge invariant operators, induced by (color, hyper-, weak) charged new vector-like fermions \cite{Lykken,whatistheresonance,DiChiara,Falkowski,Gersdorff}. We will not specify these new states until the last section. After electroweak symmetry breaking couplings with the $\gamma\gamma$, $ZZ$, $Z\gamma$ and $W^{+}W^{-}$ are induced. It was found that the production rate in the four final states depends only on two parameters $\kappa_{B}$ and $\kappa_{W}$ providing predictions for their ratios \cite{Lykken,Dev:2015vjd,Belyaev:2015hgo,McDermott:2015sck}.

 The study of the elastic $2\rightarrow2$ processes was important to find the limits and successors of the Fermi four-fermion interactions. Requiring perturbative unitarity the charged and neutral gauge bosons and finally the Higgs have to be introduced to tame the amplitudes that grow with energy \cite{BenLee}, it was mentioned in connection with the resonance in \cite{Fabbrichesi:2016alj}. In this letter the effective interactions induce derivative couplings of the new resonance to the $W^{\pm}$, $Z$, $\gamma$ leading to non-trivial two particle gauge boson scatterings growing badly with  energy $\sqrt{s}$. We find the perturbative limits of the effective theory describing the the diphoton excess, where new perturbative physics and/or strong interactions must enter. The production mechanism of the resonance in proton-proton collisions is mostly accepted to be either gluon-gluon  or $\gamma\gamma$ fusion. The allowed range of the couplings was given in several papers \cite{DiChiara,Low:2015qep,Lykken,Alves:2015jgx,Altmannshofer:2015xfo,Ellis:2015oso,Petersson:2015mkr,Bardhan:2016rsb,Cao:2015pto,Kobakhidze:2015ldh,Fichet:2015vvy,Han:2015dlp,Feng:2015wil,Heckman:2015kqk,Berthier:2015vbb,Craig:2015lra,Stolarski:2016dpa,Gersdorff,Buttazzo:2015txu,Kamenik:2016tuv,Harigaya:2015ezk}. Here we will show that the couplings are large and using perturbative unitarity we get rather low upper bounds on the validity of the effective model, nearly ruling out the large width scenario and disfavoring the production by light by light scattering  \cite{Csaki:2015vek}.

The rest of the paper is organized as follows. In the next section we give the effective interactions and calculate the two-particle elastic scattering processes for gauge bosons. We start with the $\gamma$ asymptotic states as the $\gamma\gamma$ interactions were directly observed by the two LHC experiments. Then with reasonable assumptions on the couplings we consider the massive gauge bosons scatterings and establish constraints. In section 3 we resolve the effective coupling with new charged heavy fermions, prove our assumptions about the relation of the  $\kappa_{B}$ and $\kappa_{W}$ couplings and set limits on the mass of the new fermions running in the loops. The paper is closed with conclusion and comments on the literature.

\section{VV scattering via S exchange}
The interactions of an electroweak singlet scalar $S$  with the vector bosons first induced at one-loop level. In this section, we study the effective model of the $S$ resonance.

We will follow the notation of \cite{Lykken}, the five-dimensional gauge invariant operators are
\begin{equation}
	\frac{\alpha_{em}}{4\pi s_w^2}\frac{\kappa_W}{4m_S}SW^a_{\mu\nu}W^{a\mu\nu} + \frac{\alpha_{em}}{4\pi c_w^2}\frac{\kappa_B}{4m_S}SB_{\mu\nu}B^{\mu\nu} + \frac{\alpha_s}{4\pi}\frac{\kappa_g}{4m_S}SG^a_{\mu\nu}G^{a\mu\nu} .
    \label{dim5}
\end{equation} 
The Standard Model couplings are explicitly taken out as they would appear in the one-loop integrated renormalizable model.
After electroweak symmetry breaking, the couplings to the $\gamma,Z$ bosons are 
\begin{equation}
	\Gamma_{SV_1V_2}^{\mu\nu} = \frac{\kappa_V}{m_S}\frac{\alpha_{em}}{4\pi} \left( p_{V_1} \cdot p_{V_2} g^{\mu\nu} - p_{V_1}^\nu p_{V_2}^\mu \right),
\end{equation}
with $V=\gamma\textrm{ or }Z$ and $\kappa_\gamma = \kappa_B + \kappa_W$, $\kappa_Z = \frac{c_W^2}{s_W^2}\kappa_W + \frac{s_w^2}{c_w^2}\kappa_B$. The coupling to $W^\pm$ is
\begin{equation}
	\Gamma_{SW_1W_2}^{\mu\nu} = \frac{\kappa_W}{m_S}\frac{\alpha_{em}}{4\pi s_w^2} \left( p_{W_1} \cdot p_{W_2} g^{\mu\nu} - p_{W_1}^\nu p_{W_2}^\mu \right).
\end{equation}

In the case of a pseudo-scalar resonance $S$, both the calculations and conclusions are effectively the same in this section \cite{Lykken,Molinaro:2015cwg}. 
Investigating the elastic vector boson scattering processes, we set bounds on the validity of the effective description.

\subsection*{$\gamma\gamma$ scattering} 
First, we calculate the $\gamma\gamma$ scattering, since the only experimentally observed decay channel is $S\to\gamma\gamma$. We consider only the scattering via $S$ exchange that gives amplitudes growing with $s$. The Standard Model contributions are small and  can be neglected. The three relevant Feynman graphs are shown in figure \ref{fig:feyngraphs}. We chose a concrete basis for the transverse polarization vectors, included all helicity channels in that basis for the amplitude,
\begin{multline}
	i\mathcal{M}_{\lambda_1\lambda_2\to\lambda_3\lambda_4} = \frac{i}{8m_S^2}\left( \frac{\alpha_{em}\kappa_\gamma}{4\pi} \right)^2 \\
    \cdot\begin{pmatrix}
		\frac{s^2}{s-m_S^2} + \frac{t^2\cos^2{\theta}}{t-m_S^2} + \frac{u^2\cos^2{\theta}}{u-m_S^2} & 0 & 0 & \frac{s^2}{s-m_S^2} \\
        0 & \frac{t^2\cos{\theta}}{t-m_S^2} & \frac{u^2\cos{\theta}}{u-m_S^2} & 0 \\
        0 & \frac{u^2\cos{\theta}}{u-m_S^2} & \frac{t^2\cos{\theta}}{t-m_S^2} & 0 \\
        \frac{s^2}{s-m_S^2} & 0 & 0 & \frac{s^2}{s-m_S^2} + \frac{t^2}{t-m_S^2} + \frac{u^2}{u-m_S^2}
	\end{pmatrix}.
\end{multline}
We calculate the s-wave partial amplitude and impose partial wave unitarity on the helicity channels \cite{BenLee},  
\begin{equation}
	a_0 = \frac{1}{32\pi}\int_{-1}^1 d\left( \cos{\theta} \right) \mathcal{M},
\end{equation}
\begin{equation}
	|\textrm{Re}a_0| \leq \frac{1}{2}.
\end{equation}
The best bound comes from a helicity changing channel, where only the s-channel is present, $\mathcal{M}_{\pm\pm\to\mp\mp}\propto \frac{s^2}{s-m_S^2}$,
\begin{equation}
	\sqrt{s} \lesssim  \frac{32\pi^\frac{3}{2} m_S}{\alpha_{em}\kappa_\gamma} = \frac{1.7\times 10^7 \textrm{ GeV}}{\kappa_{\gamma}}.
\end{equation}

Taking the photon coupling from \cite{whatistheresonance}, $\kappa_\gamma \in [23.7,143.1] \, (1012.0)$, where the range is from the minimal total width $\Gamma_{\textrm{tot}}=\Gamma(S\to\gamma\gamma)+\Gamma(S\to gg)$ with the lower $\kappa$ values provided by production dominated by gluon fusion and the higher values by photon fusion\footnote{This range is in agreement with other estimations in the literature. In \cite{Gersdorff}, the range of the couplings is the same, in their notation $\frac{1}{f_B}=\frac{\alpha_{em}}{4\pi c_w^2}\frac{\kappa_B}{4m_S}$ and $\frac{1}{f_W}=\frac{\alpha_W}{4\pi s_w^2}\frac{\kappa_B}{4m_S}$. \cite{Falkowski} only considers  gluon fusion dominated production  and takes large QCD $k$-factors into account, in their notation $c_{S\gamma\gamma} = \frac{v}{m_S}\frac{\kappa_\gamma}{16\pi^2}$, where $v$ is the Higgs vev.}. The preferred one is $\Gamma(S\to gg)\gg\Gamma(S\to\gamma\gamma)$ from comparing the 8 and 13 TeV LHC data. The third value given in parenthesis is coming from the best fit of ATLAS with large $S$ width. 
\begin{equation}
	\sqrt{s} \lesssim 118.8\dots 716.3 \textrm{ TeV} \, (16.8\textrm{ TeV})
\end{equation}
The lower bound  comes from the photon fusion dominated production of $S$, while at the upper end the gluon fusion dominates. As the photon has only transverse polarization, the amplitudes only grow with $s$, giving these relatively weak, but important bounds as they are based on the only experimentally observed decay channel without further assumptions. 

The calculation of gluon-gluon scattering goes similarly, but even less constrained by the experimental data \cite{whatistheresonance}. Stronger bounds are expected from massive gauge boson scatterings, which have longitudinal polarization, but the couplings are less contrained.

\subsection*{$W^+W^-$ and $ZZ$ scatterings} 
The longitudinal polarization of the massive $W^\pm,Z$, $\epsilon_L^\mu (k) \approx \frac{k^\mu}{m_{W,Z}}$ gives $2 \to 2$ amplitudes  growing with $s^3$, resulting in a strong bound on the validity of the theory. We can also investigate the $\gamma Z$ scattering as well to get slightly better bounds than from $\gamma\gamma$, but not as strong as from $ZZ$ only and with the same uncertainty. Inclusion of the Higgs in the scattering processes needs assumptions about further effective operators, not preferred by experimental observations.

\begin{figure}
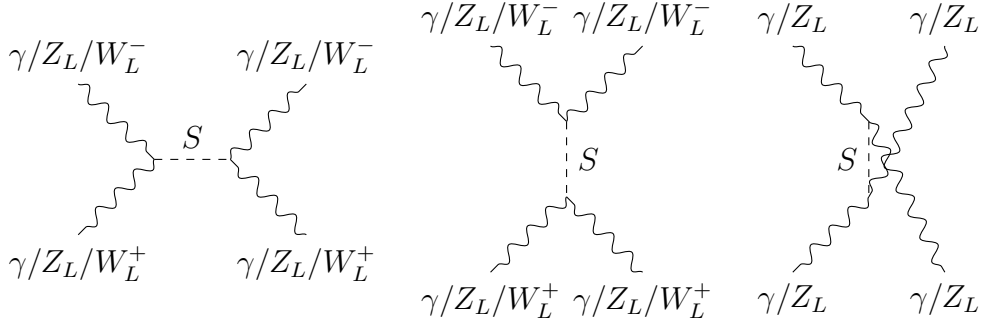
 
	\centering
	\vvsvvS{{\gamma/Z_L/W_L^-}}{{\gamma/Z_L/W_L^+}}{{\gamma/Z_L/W_L^-}}{{\gamma/Z_L/W_L^+}}{S}
    \vvsvvT{{\gamma/Z_L/W_L^-}}{{\gamma/Z_L/W_L^+}}{{\gamma/Z_L/W_L^-}}{{\gamma/Z_L/W_L^+}}{S}
    \vvsvvU{{\gamma/Z_L}}{{\gamma/Z_L}}{{\gamma/Z_L}}{{\gamma/Z_L}}{S}
    \caption{Feynman graphs for the $VV\to S\to VV$ scatterings.} \label{fig:feyngraphs}
\end{figure}

The related two and three Feynman graphs are shown in figure \ref{fig:feyngraphs} and the scattering amplitudes are the following,
\begin{equation}
	i\mathcal{M}_{W_LW_L\to W_LW_L} = \frac{i}{32m_S^2m_W^4}\left(\frac{\alpha_{em}\kappa_W}{4\pi s_w^2}\right)^2 \left( \frac{s^2(s-4m_W^2)^2}{s-m_S^2}+\frac{t^2(t-m_W^2)^2}{t-m_S^2} \right),
\end{equation}
\begin{equation}
	i\mathcal{M}_{Z_LZ_L\to Z_LZ_L} = \frac{i}{32m_S^2m_Z^4}\left(\frac{\alpha_{em}\kappa_Z}{4\pi}\right)^2 \left( \frac{s^2(s-4m_Z^2)^2}{s-m_S^2}+\frac{t^2(t-4m_Z^2)^2}{t-m_S^2}+\frac{u^2(u-4m_Z^2)}{u-m_S^2} \right).
\end{equation}
Imposing the partial wave unitarity on the amplitudes  we get for the $W^\pm$ and $Z$ scattering at leading order in $\frac{m_{V}^2}{s} $, respectively
\begin{equation} \label{eq:Wbound}
	\sqrt{s}\lesssim 4\sqrt{\pi}\sqrt[3]{\frac{2m_W^2 m_S s_w^2}{\sqrt{3}\alpha_{em}\kappa_W}} = \frac{3.86\textrm{ TeV}}{\sqrt[3]{\kappa_W}},
\end{equation}
\begin{equation} \label{eq:Zbound}
	\sqrt{s} \lesssim 4\sqrt{\pi}\sqrt[3]{\frac{2m_Z^2 m_S}{\alpha_{em}\kappa_Z}} = \frac{7.35\textrm{ TeV}}{\sqrt[3]{\kappa_Z}}.
\end{equation}
The next to leading order calculations only differ at the percent level. The bound from the $W^+W^-$ scattering is stronger, but when $\kappa_W$ vanishes, the $ZZ$ scattering bound becomes important. 

To quantify the constraints in \eqref{eq:Wbound} and \eqref{eq:Zbound}, we need more information on the couplings, $\kappa_W$ and $\kappa_B$. The experiments so far only constrains $\kappa_\gamma=\kappa_W+\kappa_B$. 

In general, the $\kappa$ couplings have the same sign\footnote{It is true if there are only fermions with same sign Yukawas and no scalars involved in resolving the effective couplings, see section \ref{sec:fermionloop}.}, assuming they are positive gives a lower and an upper bound on the $\kappa_B$ and $\kappa_W$. 
\begin{equation}
	0 \leq \kappa_{W,B} \leq \kappa_W + \kappa_B = 23.7\dots 143.1 \, (1012.0)
\end{equation}
Purely mathematically translating this to the effective $Z$ coupling $\kappa_Z = \frac{c_w^2}{s_w^2}\kappa_W + \frac{s_w^2}{c_w^2}\kappa_B \in [7.1,479.2]\, (3388.1)$. Using the value of $\kappa_Z$ in \eqref{eq:Zbound} gives a stringent limit on the validity of the effective description. Let us specify the constraints in the two limiting case of vanishing $\kappa_W$ or $\kappa_B$. 

When $\kappa_W = 0$, the $Z$ coupling is small, too, $\kappa_Z \in [7.1,42.8]\, (302.3)$, giving the energy bound
\begin{equation}
	\sqrt{s} \lesssim 2.1\dots 3.8 \textrm{ TeV} \, (1.1\textrm{ TeV}).
\end{equation}
Again, the lower end corresponds to photon fusion being the dominant production channel, while the upper bound comes from production by gluon fusion. 

In the other limit when $\kappa_B = 0$, the $Z$ coupling is much larger, $\kappa_Z\in [79.5,479.2]\, (3388.1)$, and the energy bound becomes $\sqrt{s}\lesssim 0.9\dots 1.7 \textrm{ TeV} \, (0.5\textrm{ TeV})$. In this case, the $W^+W^-$ scattering provides an even stronger bound from \eqref{eq:Wbound} with $\kappa_W = \kappa_\gamma \in [23.7,143.1]\, (1012.0)$,
\begin{equation}
	\sqrt{s}\lesssim 0.7\dots 1.3 \textrm{ TeV} \, (0.4\textrm{ TeV}).
    \label{eq:Wc}
\end{equation}
As we can see, the large width scenario gives extremely small upper bounds given in the parenthesis that are smaller than the mass of the resonance $m_S=750\textrm{ GeV}$.
We arrived at these bound with assumptions that one of the two couplings $\kappa_W, \kappa_B$  vanishes. As we can see in the section,  $\kappa_W$ and $ \kappa_B$  are expected to be the same order of magnitude, as they are generally induced at one-loop by particles having ordinary weak charges and hypercharges. For a standard vector-like doublet $\kappa_W= \kappa_B$ and we get similar bounds as in \eqref{eq:Wc}.

\section{Heavy fermion loop} \label{sec:fermionloop} 

Let us consider the case, where the above effective theory is coming from a UV complete renormalizable theory with a new heavy fermion $T$ with mass $m_T$ that couples to $S$ via a Yukawa type coupling. The mass $m_T\gtrsim 375$ GeV to avoid the unobserved direct decay of $S$ to charged fermions further leading to larger total width and large $\Gamma(S\rightarrow \gamma \gamma)$. The new operators are then
\begin{equation}
	\bar{T}i\gamma^\mu D_\mu T -\lambda_{ST} \bar{T}TS.
\end{equation}

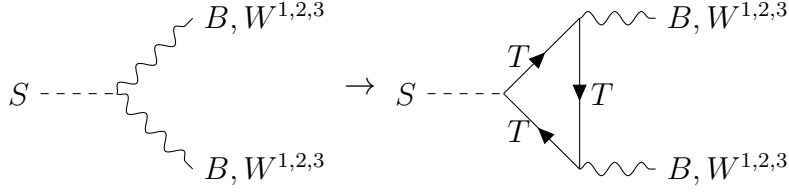
\begin{figure}
\begin{center}
\begin{tikzpicture}[baseline=(current bounding box.center)]
	\coordinate[label=left:$S$] (e1);
    \coordinate[right=of e1] (i1);
    \coordinate[above right=of i1,label=right:${B,W^{1,2,3}}$] (e2);
    \coordinate[below right=of i1,label=right:${B,W^{1,2,3}}$] (e3);
    \draw[scalar] (e1) -- (i1);
    \draw[vboson] (i1) -- (e2);
    \draw[vboson] (i1) -- (e3);
\end{tikzpicture}
 $\rightarrow$
\begin{tikzpicture}[baseline=(current bounding box.center)]
	\coordinate[label=left:$S$] (e1);
	\coordinate[right=of e1] (i1);
    \coordinate[above right=of i1] (i2);
    \coordinate[below right=of i1] (i3);
    \coordinate[right=of i2,label=right:${B,W^{1,2,3}}$] (e2);
    \coordinate[right=of i3,label=right:${B,W^{1,2,3}}$] (e3);
    \draw[scalar] (e1) -- (i1);
    \draw[fermion] (i1) -- node[label=left:$T$]{} (i2);
    \draw[fermion] (i2) -- node[label=right:$T$]{} (i3);
    \draw[fermion] (i3) -- node[label=left:$T$]{} (i1);
    \draw[vboson] (i2) -- (e2);
    \draw[vboson] (i3) -- (e3);
\end{tikzpicture}
\caption{Resolving the effective $SVV$ couplings with heavy fermion loop.}
\end{center}
\end{figure}

Calculating the couplings to the $SU(2)_W$ and $U(1)_Y$ generators gives $\kappa_W$ and $\kappa_B$  directly, see Fig. 2. This calculation justifies the separation of the electroweak couplings and loop factors in \eqref{dim5}. It can be seen that the scale of the underlying physics is $m_T$ instead of $m_S$, though they are partly related, but $m_T$ is unknown at the level of the effective theory. For a $T$ with $N_C$ colors and $N_F$ flavors, 
\begin{equation}
	\frac{\kappa_W}{2m_S} = \frac{\lambda_{ST}}{m_T}N_CN_F\sum_T T_{3T}^2 \tilde{f}(\tau),
    \label{eq:kw}
\end{equation}
\begin{equation}
	\frac{\kappa_B}{2m_S} = \frac{\lambda_{ST}}{m_T}N_CN_F\sum_T Y_{T}^2 \tilde{f}(\tau).
    \label{eq:kb}
\end{equation}
where $\tilde{f}(\tau) = \tau (1+(1-\tau)f(\tau))$ with 
\begin{equation}
	f(\tau) = \begin{cases}
		\left( \arcsin{\frac{1}{\sqrt{\tau}}} \right)^2 & \tau > 1 ,\\
        -\frac{1}{4} \left( \log{\frac{1+\sqrt{1-\tau}}{1-\sqrt{1-\tau}}} -i\pi \right)^2 & \tau < 1 ,
	\end{cases}
\end{equation}
where $\tau = \frac{4m_T^2}{m_S^2}$ and the relevant region is $\tau > 1$. There $1 > \tilde{f}(\tau) > \frac{2}{3}$ and for $m_T\gtrsim m_S$, $\tilde{f}(\tau) \approx \frac{2}{3}$ is a good approximation. That means that the ratio of the effective couplings are determined only by the $SU(2)_W \times U(1)_Y$ charges of the new fermion. 
\begin{equation}
	\frac{\kappa_W}{\kappa_B} = \frac{\sum_T T_{3T}^2}{\sum_T Y_T^2}
\end{equation}

From  Eqs. \eqref{eq:kw} and \eqref{eq:kb} we see that $\kappa_W$ and $\kappa_B$ have the same sign, depending on the common Yukawa coupling. We can imagine opposite sign  $\kappa_W$, $\kappa_B$ with more fermions in different representations and Yukawas or additional scalars in the loop.

Let us now consider a concrete model with the new fermion $T$ is a vector-like $SU(2)_W$ doublet without color charge, $T=\begin{pmatrix} T^0 \\ T^- \end{pmatrix}$, with the quantum numbers $T_{3T}=\frac{1}{2}$ and $Y_T=\frac{1}{2}$. Then from the fermion loops, the ratio of the couplings $\frac{\kappa_B}{\kappa_W} = 1$. When these two couplings are equal $\kappa_W = \kappa_B = \frac{\kappa_\gamma}{2} \in [11.9,71.6]\, (506.0)$, the strongest bounds come from the $W^+W^-$ scattering.  
\begin{equation}
	\sqrt{s} \lesssim 0.9\dots 1.7 \textrm{ TeV} \, (0.5\textrm{ TeV}).
    \label{eq:dubc}
\end{equation}
The bounds are similar to \eqref{eq:Wc}, the $\kappa_B=0$  case.
The ATLAS large width scenario leaves nearly no room for the effective interactions in \eqref{dim5}, as $m_T\gtrsim 375\textrm{ GeV}$. The lower values in \eqref{eq:Wc} and \eqref{eq:dubc} coming from the photon fusion production give rather low bounds, this way making the  gluon fusion dominated process with minimal total width the most likely scenario.

In this case, we can also bound the ratio of the new fermion's Yukawa coupling and mass from $\kappa_W$,
\begin{equation}
	\frac{\lambda_{ST}}{m_T} = \frac{\kappa_W}{2m_S}\frac{1}{\tilde{f}(\tau)} \overset{\tau>1}{\rightarrow} \frac{3\kappa_W}{4m_S} \in [0.008,0.05]\,(0.34).
\end{equation}
\begin{equation}
	m_T \approx \lambda_{ST}\cdot (21 \dots 126\textrm{ GeV}) \quad (\lambda_{ST} \cdot (3\textrm{ GeV})).
\end{equation}
The perturbative limit on the  Yukawa coupling is $\lambda_{ST}\lesssim 4\sqrt{\pi}$, where the previous mass limits translate to
\begin{equation}
	m_T \approx 148 \dots 896\textrm{ GeV} \quad (21\textrm{ GeV}).
    \label{eq:mtc}
\end{equation}
These low bounds are valid if $T$ is a color-singlet and the $TS$ Yukawa coupling is close to its perturbative limit, $\lambda_{ST} \approx 4\sqrt{\pi}$. The  perturbative unitarity bound for the resolved interactions, e.g. for  weakly charged vector-like fermion doublets were calculated in \cite{juccus} giving similar constraints. To have smaller Yukawa couplings, we can take fermions in color (or flavor) multiplets, that gives a factor of $N_C$ (or $N_F$), allowing $\mathcal{O}(1)$ Yukawa couplings. Several fermion multiplets can be considered to ease the strong bounds. It is clear from \eqref{eq:mtc} that for a single fermion the gluon dominated production is favored, the $\gamma \gamma $ fusion and the large width scenario are ruled out. For the large width a huge multiplicator $N_C N_F$ is needed to reach the  S decay threshold 375 GeV. With colored fermions in the loop the gluon coupling is similarly induced as \eqref{eq:kw} and \eqref{eq:kb}, but we did not need it in our analysis. For the  photon-photon fusion large $N_C \cdot N_F$ is needed, at least a colored $T$ quark is favored and possibly in more than one generation to have Yukawa couplings not saturating the perturbativity bound.

\section{Conclusion} 

We studied the effective model for the newly postulated 750 GeV resonance $S$ with perturbative unitarity. We have considered the one-loop generated dimension five interactions where the loop and gauge coupling factors were separated. Then calculated the two-particle elastic scattering amplitudes for various final states and imposed the perturbative unitarity on $a_0$ partial wave amplitude. Even though the experimental bounds are only available for the $S\gamma\gamma$ coupling, with requiring positivity for the $\kappa$ couplings, the effective theory is limited to be valid below $\sqrt{s}\lesssim 1.3 \dots 3.8\textrm{ TeV}$. This is the scale where new degrees of freedom or strongly interacting dynamics responsible for the resonance should appear.  From not observing excess in the other $VV$ channels we could further bound the effective $\kappa_W$, $\kappa_Z$ couplings. Moreover, the $SZZ$ and $SW^+W^-$ couplings can be generated independently from  the $S\gamma\gamma$ coupling by other effective operators, such as $\frac{1}{f_H}|D_\mu H|^2$, see \cite{Gersdorff,Kamenik:2016tuv}.

If the one-loop induced couplings are expected to be equal, which is the case of a vector-like weak doublet  with half hypercharge running in the loop, we get strong limits $\sqrt{s}\lesssim 0.9\dots 1.7\textrm{ TeV}$ for the minimal total width scenario. For the large $S$ width preferred by ATLAS the onset of new physics is at approximately 0.5 TeV, extremely low. This constrains the mass of the vector-like fermion $m_T$ between $375\dots 500 \textrm{ GeV}$, which still requires large Yukawa coupling turning the scenario unlikely. The lower bound in \eqref{eq:dubc} is just above the mass of the resonance, implies large $\kappa$ and Yukawa couplings disfavor the $\gamma \gamma$ fusion dominated production. This is because the large Yukawas soon develop a Landau pole during the renormalization group running. The Landau pole can be avoided by adding enough number of new fermions, which requires fine tuning to circumvent the instability of the scalar potential \cite{Son,planck}. The resolution of the effective couplings presented in section 3 also points towards colored and high multiplicity fermions in the loop to avoid large Yukawas. All these findings imply that if the LHC experiments prove the existence of the postulated resonance, its perturbative treatment prefers production by gluon fusion and the relatively early onset of new physics resolving the effective interactions. A second solution can be a similarly early starting strong dynamics \cite{Son,Fodor,Bian:2015kjt,Sannino:2004qp,Dietrich:2006cm} that should show up as new resonances at the LHC.

\end{document}